\documentclass[11pt, a4paper]{article}

\usepackage{amsmath,amssymb}
\usepackage{graphicx}
\usepackage{float}                  
\usepackage{booktabs}               
\usepackage{siunitx}                
\usepackage[margin=1.0in]{geometry} 
\usepackage{easyReview}             
\usepackage{hyperref}              
\usepackage{authblk}

\title{Performance of the modified Becke-Johnson potential employing the pseudopotential plane-wave approach for band structure calculations}
\author[1]{Hazem Abu-Farsakh}
\author[2]{Abdallah Qteish}
\affil[1]{Department of Mathematics and Sciences, Prince Sultan University, Riyadh 11586, Saudi Arabia}
\affil[2]{Department of Physics, Yarmouk University, Irbid 21163, Jordan}
\providecommand{\keywords}[1]{\textbf{\textit{Keywords:}} #1}
\date{}
\begin{document}
\maketitle
\begin{abstract} 
	The modified Becke-Johnson exchange potential combined with local-density approximation 
	correlation (mBJLDA) has recently attracted interest because it provides highly improved 
	band gaps at a very low computational cost. In this work we performed an extensive 
	investigation of the performance of the mBJLDA potential employing a norm-conserving 
	pseudopotential plane-waves approach (mBJLDA@PP), as implemented in the \textsc{abinit} 
	code, using a test set of 83 solids representing a wide range of semiconductors and 
	insulators. Our results confirm the conclusion of our previous study that the number of 
	electrons treated as valence in the pseudopotentials of the cations can have a significant
	impact on the calculated mBJLDA@PP band gaps. More specifically, while the use of typical 
	pseudopotentials leads to accurate band gaps of certain systems, it yields significantly 
	underestimated band gaps for other systems compared to experiment and to those of the 
	all-electron mBJLDA (mBJLDA@AE) approach. The 
	classes of the latter systems are identified, and this problem is resolved by including some 
	outer core states as valence. The resulting mean absolute error in the 
	calculated band gaps (compared to experiment) is of 0.46 eV, which is comparable to that 
	of the mBJLDA@AE band gaps, reflecting the accuracy and reliability of the mBJLDA@PP approach 
	for the band gap calculations.
\end{abstract}
\noindent\keywords{electronic structure, band gap, semiconductors, insulators, density functional theory, mBJLDA
	potential, pseudopotentials}

\section{Introduction}
The Kohn-Sham formalism of the density functional theory (KS-DFT) 
\cite{Kohn1965,Hohenberg1964} provides an efficient approach for calculating 
materials properties. However, the required approximation for the exchange and 
correlation (XC) energy component of the effective (known as KS) potential imposes 
some limitations. For example, the widely used local density (LDA) and the generalized 
gradient (GGA) approximations are known to significantly underestimate the band gaps of semiconductors 
and insulators, which is a key quantity for many technological applications. It is worth to 
mention that the KS-DFT formalism is a ground-state method, and hence it is not supposed to 
provide accurate band gaps due to a derivative discontinuity in the XC potential \cite{Godby1988}. 
On the other hand, the GW approach \cite{Rinke2005} provides a formal way for 
calculating the band gaps but at a significantly higher computational cost, which 
makes GW calculations for large systems prohibitive. 

In order to provide an improved description of the band gaps within DFT, several 
approaches for approximating the XC potential have been devised, such as semilocal 
functionals (see for example Refs. \cite{Borlido2020, Borlido2020a, Yang2016}) and 
hybrid-functionals \cite{Becke1993,Heyd2003}. Of particular interest here is the modified 
Becke-Johnson \cite{Becke2006} exchange-potential combined with LDA correlation (mBJLDA) 
\cite{Tran2009}, which has gained popularity over the past decade due to its 
ability to provide band gaps with an accuracy comparable to that of the more 
advanced GW and hybrid functionals approaches \cite{Tran2019}, at a much lower computational 
costs. The mBJLDA potential has been extensively tested on a wide range of materials, 
such as the 40-semiconductor test set (SC40) \cite{Ye2015}, the test set of 76 
solids \cite{Tran2017}, and 472 materials \cite{Borlido2019}. 
The success of the mBJLDA potential to provide highly improved band gaps is attributed to the 
contribution of the kinetic energy density term, which affects valence and conduction electrons 
differently, leading to a band gap opening \cite{Koller2011,Abu-Farsakh2021}. Despite the 
impressive successes of the mBJLDA potential, it is worth to mention that it does not provide 
significantly improved band gaps for some materials such as Cu$_2$O \cite{Koller2011} and 
ScF$_3$ \cite{Hamed2015}. Moreover, mBJLDA is found \cite{Abu-Farsakh2021,Kim2010} to 
overestimate effective masses, to underestimate bandwidths, and to underestimate binding 
energies of semicore states. This is not surprising because the mBJLDA potential is optimized 
to reproduce the experimental band gaps \cite{Tran2009}.

The pseudopotential plane-waves (PP-PW's) approach is widely used and has proved 
to provide a comparable accuracy to all-electron methods \cite{Borlido2020}. It is 
worth noting that the mBJLDA potential was optimized using full-potential linearized 
augmented plane-waves (FP-LAPW) method \cite{Tran2009}, which is an all electron 
(AE) approach, and hence it is not, in principle, supposed in to provide a good description 
of the band gaps when used in conjunction with the pseudopotential approach. In spite of 
this fact,  it has been implemented in several pseudopotential plane waves codes, such as 
\textsc{abinit} \cite{Gonze2020,Romero2020}, Quantum Espresso \cite{Germaneau2013}, 
and \textsc{castep} \cite{Bartok2019}. Very recently, we have thoroughly investigated 
\cite{Abu-Farsakh2021} the performance of the mBJLDA pseudopotential (mBJLDA@PP) 
approach, as implemented in \textsc{abinit}, for electronic structure calculations 
(band gaps, effective masses, band widths, and binding energies of semicore electrons), using 
11 carefully selected systems. We found that the improvement in the 
mBJLDA@PP band gaps using typical pseudopotentials is not systematic. For some systems the 
so-calculated band gaps are significantly underestimated compared to those of the
all-electron mBJLDA (mBJLDA@AE) ones, while accurate band gaps are obtained for the 
other considered systems. 
Interestingly, we have shown that the band gaps of the former systems can be highly
improved if the uppermost $p$ states are included as valence in the cation PP's. 
The so-obtained band gaps and the other considered electronic 
structure properties (see above) are found to be in very good agreement with the 
corresponding mBJLDA@AE results. Therefore, in accordance with the mBJLDA@AE approach, only 
the band gap is well descried by the mBJLDA@PP method \cite{Abu-Farsakh2021,Kim2010}. 

The main objective in this work is to further investigate the effects of the number of 
electrons treated as valence (NETV) in the PP's of the cations on the mBJLDA@PP band gaps 
using a large set of test systems. The considered set contains 83 solids, 
which combines the test set of 76 solids used in Ref. \cite{Tran2017} and the 
40-semiconductors (SC40) test set of simple and binary semiconductors used in Refs. 
\cite{Heyd2005,Ye2015} (see Table S1 in the Supporting Information). These solids 
represent a wide range of semiconductors and insulators. The results obtained using 
different NETV will be discussed in comparison with the already available mBJLDA@AE 
band gaps \cite{Tran2017, Borlido2019, Ye2015} and experimental data \cite{Abu-Farsakh2021}.

\section{Computational details}
\label{Section:comp_details}

As in our previous work \cite{Abu-Farsakh2021}, the LDA and mBJLDA@PP calculations were 
performed using the \textsc{abinit} code \cite{Gonze2020,Romero2020} and optimized 
norm-conserving Vanderbilt pseudopotentials \cite{Hamann2013}. The generation of the 
PP's is also performed as described in Ref. \cite{Abu-Farsakh2021}. To investigate the 
effect of the NETV in the PP's of the cations, we have classified the PP's according to the used 
NETV to four main categories: 
(i) typical pseudopotentials (t-PP's), in which valence and semicore states are 
pseudized. The adopted PP's of the anions are of this type. 
(ii) PP's in which the outer core $p$ electrons are additionally treated as valence, 
referred to as cp-PP's. 
(iii) PP's in which the outer core $s$ and $p$ electrons are included in the 
valence, referred to as csp-PP's. 
(iv) In this work, we encountered a new situation in which the uppermost core states are of 
$d$ character. This is the case for Sr, Mo, and Ba atoms
(see Table S2 in the Supporting Information), which belong to the 5\textsuperscript{th} 
and 6\textsuperscript{th} periods of the Periodic Table.  For these 
elements, this fourth category is therefore considered, in which the upper core $d$ electrons 
are treated as valence, hereafter referred to as cd-PP's. It is worth noting that for these 
three elements the outer core $d$ electrons are also implicitly included as valence in their 
so-called cp- and csp-PP's. 

The other computational details are as described in Ref. \cite{Abu-Farsakh2021}. We note 
that currently the \textsc{abinit} code does not support PP's with non-linear core 
correction (NLCC) for meta-GGA's functionals, and therefore the NLCC was not included. 
All the generated PP's were checked to be free of ghost states. Convergence tests with 
respect to the cutoff energies for the PW's basis were performed for each system and 
for each set of PP's. The used cutoff energies and the Monkhorst-Pack grids  
\cite{Monkhorst1976} for Brillouin zone samplings produce total energies which are 
converged to better than 1 meV/atom. As it is commonly done in band structure 
calculations \cite{Tran2017,Tran2009, Jiang2016, Rinke2005}, 
all the calculations were performed at the experimental lattice parameters, 
which are taken from Refs. \cite{Heyd2005,Tran2017}.

\section{Results and discussion}

The calculated band gaps of the 83 semiconductors and insulators are listed in Table S1 
of the Supporting Information. The reported band gaps are obtained using LDA and mBJLDA 
functionals, and employing the considered PP's with different NETV (see Sec. 
\ref{Section:comp_details}). The corresponding mBJLDA@AE \cite{Tran2017,Tran2019,Ye2015} 
and experimental (taken from Refs. \cite{Tran2017,Tran2019,Ye2015}) values are also reported.
We note that the LDA band gaps are insensitive to NETV in the 
cations' PP's (see Table S1), and hereafter only those of the LDA@t-PP approach are thus
considered and referred to as LDA band gaps.

The comparison with experimental band gaps deserves some comments. Apart from the 
experimental uncertainties (of about 0.1 eV), notable differences in the experimental band 
gaps of some compounds are observed. These discrepancies can be attributed to 
the employed experimental technique, purity of the sample, and temperature \cite{Strehlow1973}.
On the other hand, the electron-phonon interaction is completely neglected in our and in almost 
all other mBJLDA, hybrid-functionals, and $GW$ calculations, and thus the 
comparison with experimental band gaps should be taken with caution. In the present work we 
adopted the experimental band gaps listed in Refs. \cite{Tran2017,Tran2019,Ye2015}, in 
which the mBJLDA@AE are reported.

It is worth noting that we have already shown \cite{Abu-Farsakh2021} that the mBJLDA@PP 
band gaps are sensitive to NETV in the cations PP's, which is also evident from Table S1. 
Therefore, to present and discuss our results we will classify the considered solids 
according to the group of the cation, employing the chemical abstracts service (CAS) 
numbering scheme of the Periodic Table. This will help us to 
identify trends for systems sharing cations from the same group. To analyze our results we shall focus 
on the percentage errors in the calculated mBJLDA@PP band gaps compared to experiment. 
Finally, the general features and trends will be presented and discussed. 
 
 \begin{figure}[t] 
 	\centering
 	\includegraphics[width=0.65\linewidth]{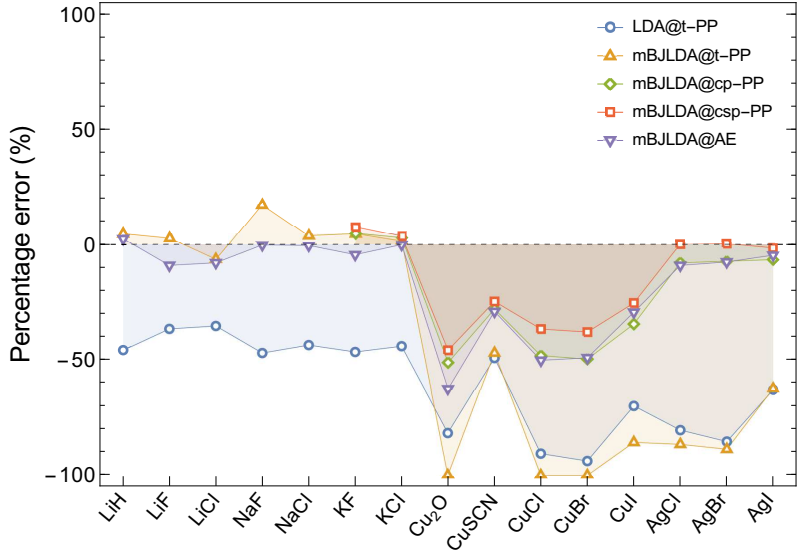}
 	\caption{Percentage error in the calculated mBJLDA@PP band gaps for group IA- and 
 		IB-based systems compared to experiment, using different sets of PP's. For 
 		comparison purposes, we also show those of the LDA@t-PP and reference 
 		mBJLDA@AE band gaps. The points are connected as a guide to the eye.}
 	\label{fig:errorsIA-IB}
 \end{figure}

\subsection{Group IA- and IB-based compounds}

The percentage errors in the calculated mBJLDA@PP band gaps of the considered IA- and 
IB-based compounds are depicted in Fig. \ref{fig:errorsIA-IB}, using the different 
sets of cation PP's (t-PP, cp-PP, csp-PP). Also shown are the percentage errors in 
mBJLDA@AE and LDA band gaps. The considered systems are mainly IA-VIIA and 
IB-VIIA binary compounds. We note that for the light cations (namely, 
Li and Na) only the t-PP's can be generated. 

We will start with the considered group IA-based compounds. The three main features 
to note from Fig. \ref{fig:errorsIA-IB} are the following: (i) The well-known 
systematic underestimation of the band gaps within the LDA approach. 
(ii) The only available mBJLDA@t-PP band gaps of the Li- and Na-based compounds are
systematically larger than the corresponding mBJLDA@AE results and the experimental 
band gaps, except for LiCl. However, it should be noted that the overall agreement with
experiment is similar to that of the mBJLDA@AE band gaps. 
(iii) For the K-based compounds (KF and KCl), including the uppermost core $2p$ or $2s+2p$ 
electrons as valence in the K pseudopotential (i.e., using respectively cp- and 
csp-PP's for K) has relatively small effects on their mBJLDA@PP band gaps -- the band gaps 
of both compounds increase slightly by increasing NETV (see Table S1). 
Therefore, one can conclude that the mBJLDA@t-PP approach provides a very good 
description of the band gaps of compounds involving group IA elements (at least for 
Li-, Na- and K-based compounds).    
 
Fig. \ref{fig:errorsIA-IB} shows that the mBJLDA@PP band gaps of the group 
IB-based compounds behave differently from the above IA-based ones. The important
features to note are the following. 
First, for the considered Cu- and Ag-based compounds, 
the mBJLDA@t-PP band gaps are significantly underestimated, and can be even in a worse 
agreement with experiment than the LDA ones. 
Secondly, including the upper core $p$ or $s+p$ electrons as part of the valence electrons 
in the Cu and Ag PP's improves significantly their mBJLDA@PP band gaps. While the best 
agreement with the mBJLDA@AE results is obtained when cp-PP's are used, it can be easily 
noted that the best agreement with experiment can be achieved when csp-PP's are used for 
both Cu and Ag. 
Thirdly, the mBJLDA@PP band gaps for Cu-based systems are still significantly underestimated, 
which is the case in both the pseudopotential and the all-electron approaches. 

For a quantitative analysis we calculate the mean absolute error (MAE) in the 
band gaps calculated by employing different approaches, compared to the experimental 
data. The MAE in mBJLDA@PP band gaps when t-PP's are used is of 1.62 eV, and it 
drops to 0.76 eV when cp-PP's are used whenever applicable\footnote{That is, when 
only t-PP's are possible then for the corresponding systems the t-PP's results 
are used. This is done so that the different MAE values always include the same number 
of systems.}. The MAE decreases further to 0.66 eV when csp-PP's are used whenever applicable.
This clearly indicates that using csp-PP's for group IB-based compounds leads to the best 
agreement with experiment. For comparison, the MAE in the corresponding mBJLDA@AE band gaps 
is of 0.67 eV. On the other hand, for the group IB-based compounds the mean absolute 
difference between mBJLDA@PP and mBJLDA@AE band gaps is of 0.26 eV (0.08 eV) 
when the csp-PP's (cp-PP's) are used, which confirms that the best agreement 
with the all-electron mBJLDA band gaps is achieved by employing the cp-PP's.

\begin{figure}[t]
	\centering
	\includegraphics[width=0.65\linewidth]{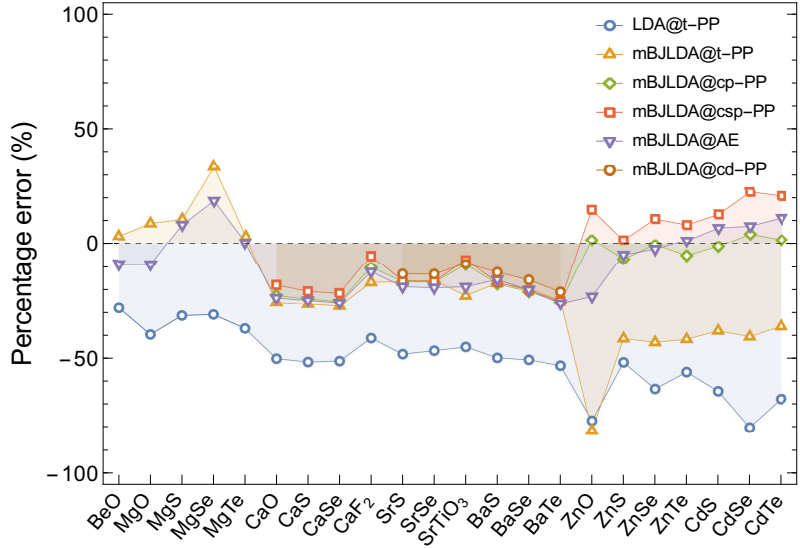}
	\caption{As in Fig. \ref{fig:errorsIA-IB} but for group IIA- and IIB-based compounds.}
	\label{fig:errorsIIA-IIB}
\end{figure}

\subsection{Group IIA- and IIB-based compounds}
\label{Section:IIA-IIB}
  
The percentage errors in the calculated mBJLDA@PP band gaps of the considered IIA- and 
IIB-based compounds, using the different sets of cations PP's (t-PP, cp-PP, csp-PP), are 
depicted in Fig. \ref{fig:errorsIIA-IIB}. Moreover, the obtained results for the Sr- and 
Ba-based compounds by using cd-PP's are also shown 
(see Sec. \ref{Section:comp_details} and Table S2 in the Supporting Information).
We note that for the light cations (namely, Be and Mg) only the t-PP's can be 
generated. We also note that CaTe and SrTe are excluded from this figure 
because of the lack of experimental band gaps, but their calculated band gaps are 
available in Table S1. Also shown are the percentage errors in mBJLDA@AE and LDA 
band gaps.

For the considered group IIA-based compounds, the important features to note from 
Fig. \ref{fig:errorsIIA-IIB} are the following. 
(i) The best agreement with the mBJLDA@AE band gaps is achieved by employing the 
mBJLDA@t-PP approach.
(ii) The best agreement with the experimental band gaps can be achieved when the
upper core electrons are treated as valence in the Ca, Sr, and Ba PP's. 
More specifically, for the considered Ca-based compounds the best agreement with 
experiment is achieved by using the mBJLDA@csp-PP approach. However, the differences 
between the mBJLDA@PP band gaps obtained using cp-PP's and using csp-PP's are relatively 
very small. On the other hand, for the Sr- and Ba-based compounds shown in 
Fig. \ref{fig:errorsIIA-IIB} the best agreement with experiment is achieved 
when cd-PP's are used, except for SrTiO$_3$. Including additional ($p$ or $s+p$) core 
electrons as valence leads to slightly smaller mBJLDA@PP band gaps (i.e., increasing 
further their underestimation). 
(iii) The mBJLDA@PP band gaps of Ca-, Sr-, and Ba-based compounds are still 
underestimated compared to experiment, which is also the case for mBJLDA@AE band gaps. 

As for group IIB-based compounds, it can be easily seen from
Fig. \ref{fig:errorsIIA-IIB} that mBJLDA@t-PP band gaps are significantly smaller than 
the corresponding mBJLDA@AE ones. This behavior is similar to that observed for group 
IB-based compounds, see above. However, contrary to group IB-based compounds, 
the mBJLDA@t-PP band gaps are better than the corresponding LDA ones, except for ZnO. 
Fig. \ref{fig:errorsIIA-IIB} also shows that the best agreement between the mBJLDA@PP
band gaps and the corresponding experimental and mBJLDA@AE band gaps can be achieved 
by using the cation cp-PP's, except for ZnS. The additional inclusion of the uppermost
core $s$ electrons (i.e., using csp-PP's) leads to overestimated band gaps compared 
to experiment. The resulting small absolute errors in mBJLDA@PP band gaps when cp-PP's
are used (between 0.02-0.26 eV) are close to the uncertainties in the experimental 
band gaps \cite{Crowley2016a}.

In summary, for this set of compounds the best overall agreement with the experimental 
band gaps is achieved when the electrons in the upper core orbital are treated as 
valence in the PP's of the cations whenever possible, except for Ca-based compounds.  
That is, by using cd-PP's for Sr and Ba, csp-PP for Ca, and cp-PP's for the remaining cations, 
except for Be and Mg where only t-PP's are possible. This can be also seen from 
the calculated MAE of the mBJLDA@PP band gaps compared to the experimental data. 
The value of the MAE when only t-PP's are used for all these systems is of 0.98 eV. 
This value decreases to 0.44 eV when the appropriate cd-, cp-, or csp-PP's are used 
(see above). On the other hand, the MAE 
increases to 0.54 eV when csp-PP's are consistently used whenever 
applicable. For comparison, the MAE of the mBJLDA@AE band gaps for these systems 
is of 0.62 eV.

\begin{figure}[t]
	\centering
	\includegraphics[width=0.65\linewidth]{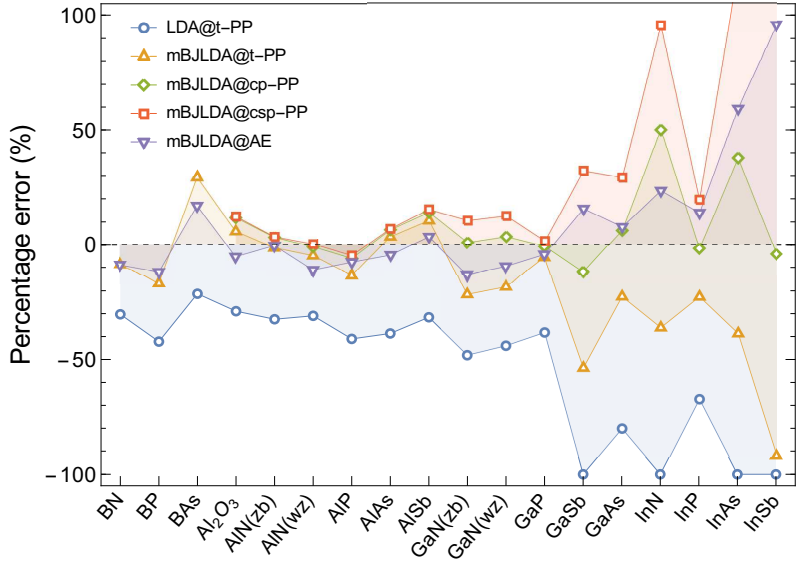}
	\caption{As in Fig. \ref{fig:errorsIA-IB} but for group IIIA-based compounds.}
	\label{fig:errorsIIIB}
\end{figure}

\subsection{Group IIIA-based compounds} 

The percentage errors in the mBJLDA@PP band gaps of the considered group IIIA-based 
compounds (using the different sets of PP's) are shown in Fig. \ref{fig:errorsIIIB}, 
together with the percentage errors in the mBJLDA@AE and LDA band gaps. We note here 
that only t-PP can be generated for B. Additionally, BSb is not included in the 
figure due to the lack of experimental band gap, and its calculated band gaps are 
listed in Table S1.

The important features to note from Fig. \ref{fig:errorsIIIB} are the following. 
(i) The mBJLDA@t-PP band gaps of the considered B- and Al-based compounds are, 
generally speaking, in a good agreement with experiment and mBJLDA@AE calculations. 
In fact, for the Al-based compounds the best agreement 
between mBJLDA@PP and mBJLDA@AE band gaps is achieved when the t-PP of Al is employed,
except for AlP.
On the other hand, using the cp- or csp-PP's of Al leads to a small further opening
of calculated mBJLDA@PP band gaps, which, in turn, leads to a slightly better 
agreement with experiment than the mBJLDA@t-PP approach for some compounds, namely AlP 
and the wurtzite phase of AlN. 
(ii) The situation is different for the considered Ga- and In-based compounds,
where large differences are observed between mBJLDA@PP band gaps obtained using 
different sets of PP's. For these systems the mBJLDA@PP band gaps are 
underestimated (overestimated) compared to the corresponding experimental data 
when t-PP's (csp-PP's) are employed. The best agreement with the experimental band gaps 
for the Ga- and In-based compounds is achieved when cp-PP's are used.

In overall, the best agreement with the experimental band gaps for group IIIA-based 
compounds with mBJLDA@PP is obtained when cp-PP's are used for the cations
whenever applicable. This is reflected in the MAE in the mBJLDA@PP band gaps, 
which decreases from 0.33 eV when only t-PP's are used to 0.22 eV when cp-PP's 
are used whenever possible, while it is of 0.36 eV when the csp-PP's are employed 
instead of the cp-PP's. We note that the MAE in mBJLDA@AE band gaps for this set 
of systems is 0.25 eV.

\begin{figure}[t]
	\centering
	\includegraphics[width=0.65\linewidth]{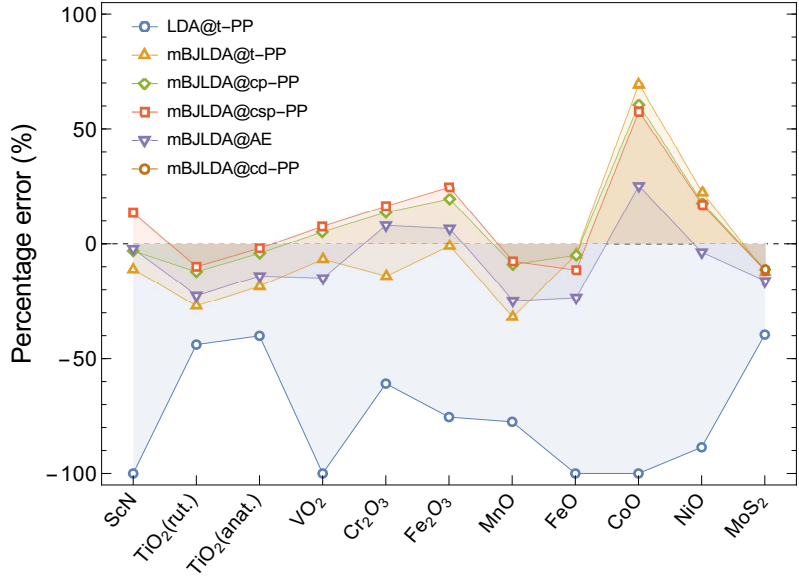}
	\caption{As in Fig. \ref{fig:errorsIA-IB} but for transition metal compounds.}
	\label{fig:errorsTM}
\end{figure}

\subsection{Transition metal compounds} 

The percentage errors in the calculated mBJLDA@PP band gaps of the considered 
transition metal compounds using the different sets of PP's are shown in 
Fig. \ref{fig:errorsTM}. This set of systems contains ScN and transition metal 
oxides, in addition to MoS$_2$. Among these systems are antiferromagnetic 
transition metal oxides (MnO, FeO, CoO, NiO, Fe$_2$O$_3$, Cr$_2$O$_3$). We note that for 
the Mo-based compounds the cd-PP of Mo is also used. Also shown in Fig. \ref{fig:errorsTM} 
are the percentage errors in calculated mBJLDA@AE and LDA band gaps. 

The three main features to note from Fig. \ref{fig:errorsTM} are the following. 
(i) For these highly correlated electron systems, LDA fails badly as expected to describe 
their band gaps: the underestimation is close to 100\% for many of them. 
(ii) The mBJLDA@t-PP approach provides a significant improvement in the calculated 
band gaps, leading generally speaking to a good agreement with the mBJLDA@AE band gaps.
This is particularly the case for systems with light cations (namely, the Sc- and
Ti-based compounds, and VO$_2$).  
(iii) Appreciable opening in the mBJLDA@PP band gaps occurs by increasing NETV, except
for FeO, CoO and NiO where small reductions take place. The mBJLDA@PP band gaps of CoO 
and NiO are overestimated, and hence increasing NETV leads to a better agreement with 
experiment. In the case of MoS$_2$ the dependence on the NETV is very weak ($\sim$0.02 eV). 
These effects lead, generally speaking, to a better agreement with experiment especially 
when the uppermost ($p$ or $d$) core electrons are included as valence in the cation PP's. 
The calculated MAE in the band gaps in this case is 0.40 eV, 
compared to 0.58 eV and 0.41 eV when t-PP's and csp-PP's are used, respectively. For 
comparison, the MAE in the mBJLDA@AE band gaps is 0.39 eV. 
(iv) The mBJLDA@PP band gaps of CoO (of 4.01 and 3.94 eV, using respectively cp- 
and csp-PPs) show the largest relative deviations from the adopted experimental value 
of 2.5 eV \cite{Elp1991}. It is worth noting that a wide range of experimental band 
gaps can be found in literature, ranging from 2.1 eV \cite{Pratt1959} to 5.43 eV 
\cite{Kang2007}. Similar variations in the reported experimental band gaps can 
be also found for MnO (2.0-4.2 eV) and NiO (3.7-4.3 eV) \cite{Crowley2016a}. 

\subsection{Group IVA-based systems, rare gas solids, and Sb$_2$Te$_3$. }

\begin{figure}[t]
	\centering
	\includegraphics[width=0.65\linewidth]{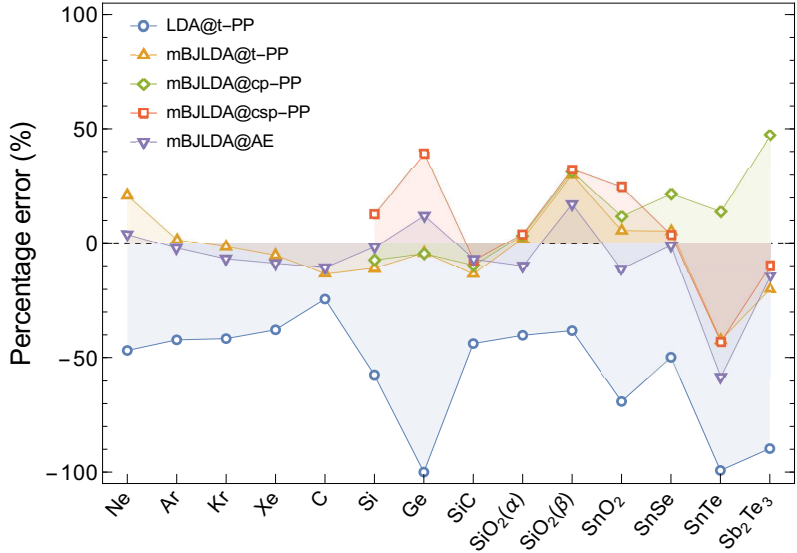}
	\caption{As in Fig. \ref{fig:errorsIA-IB} but for group IVA-based systems, 
	rare gas solids, and Sb$_2$Te$_3$.}
	\label{fig:errorsIVB}
\end{figure}

In this subsection we will present and discuss the mBJLDA@PP band gaps of the rest of 
the 83 considered systems. Namely, rare gas solids (Ne, Ar, Kr, and Xe), elemental 
group IVA solids (C, Si and Ge), some Si- and Sn-based compounds, and  
Sb$_2$Te$_3$, which is the only group VA-based compound considered here. 
The percentage errors in the calculated mBJLDA@PP band gaps using the different sets 
of PP's are shown in Fig. \ref{fig:errorsIVB}. 

It is evident from the Fig. \ref{fig:errorsIVB} that the mBJLDA@t-PP approach is 
capable of providing a very good description of the band gaps of rare gas solids 
and group-IVA elemental solids and compounds. Generally speaking, 
increasing the NETV does not lead to improved band gaps. A clear exceptional case is 
SnTe where Sn cp-PP provides a better band gap compared to experiment. Moreover, in the 
case of Sb$_2$Te$_3$ the use of Sb cp-PP leads to an overestimated band gap. We note 
that Sb$_2$Te$_3$ and SnTe have small band gaps (of 0.28 eV and 0.36 eV, respectively), and
hence small differences between the calculated and the experimental 
band gaps (e.g., of 0.13 eV and 0.05 eV respectively, when cp-PP's are used) 
lead to relatively large percentage errors.
These observations are reflected in the MAE values in 
the calculated mBJLDA@PP band gaps, which are respectively of 0.71, 0.74 and 0.79 eV 
when t-, cp- and csp-PP's are used. The MAE in mBJLDA@AE band gaps is of 0.48 eV. 
A large contribution to this difference in the MAE between the mBJLDA@AE and mBJLDA@PP 
band gaps comes from that for Ne, where the absolute error in the mBJLDA@PP band gap is 
4.52 eV, compared to 0.85 eV in the mBJLDA@AE value.

\subsection{General features and trends}

The above discussion shows that the cations of the considered systems can be classified,
according to their optimal PP's to be used in conjunction with the mBJLDA@PP approach, into 
two main categories. 
(i) Cations where t-PP's are good enough. This set of cations include groups IA and IVA 
elements, Sb, rare gases, and the light elements of groups IIA and IIIA 
(namely, Be, Mg, B and Al). Including the uppermost core states (whenever applicable) as 
valence will not significantly alter the effective potentials seen by the valence and 
conduction electrons of their considered solids, leading to only small changes in their band 
gaps by increasing NETV.
(ii) Cations where some uppermost core states have to be included as valence. This set 
of cations include transition metals, groups IB and IIB, and the remaining (heavier) 
elements of groups IIA and IIIA. For these cations including the outer core $p$ or $d$ 
(in Sr, Ba and Mo atoms) electrons as valence leads to a very good overall agreement between
the mBJLDA@PP and mBJLDA@AE band gaps. 

The above different behaviors can be attributed to the degree of overlap between the valence
and core charge densities. If such overlap is insignificant then the cation belongs to category 
(i). Otherwise, it belongs to category (ii). Including explicitly 
these core electrons as valence alters the effective potential in such a way that the 
upward shifts in the eigenvalues of uppermost valence band states are smaller than 
those of the lowest conduction band ones, leading to an opening of mBJLDA@PP band gaps. 
This is illustrated in details in Ref. \cite{Abu-Farsakh2021} for the case of ZnS. 
 
Now, focusing on the cations of type (ii), it can be easily observed (see 
Figs. \ref{fig:errorsIA-IB}, \ref{fig:errorsIIA-IIB} and \ref{fig:errorsIIIB}) that 
the opening of the band gap, by including uppermost core electrons as valence, depends 
appreciably on the group of the cation. 
Staring with group IB-based compounds, the opening of the mBJLDA@PP band gaps 
is relatively small and increases with increasing NETV. The band gaps obtained 
using both cp- and csp-PP's are smaller than the corresponding experimental band gaps 
(see Fig. \ref{fig:errorsIA-IB}), except for the Ag-based compounds. For this reason, 
the mBJLDA@PP band gaps of Ca-, Cu- and Ag-based compounds obtained by using csp-PP's 
are in better agreement with experiment than those obtained using cp-PP's. Whereas, 
the latter band gaps are in better agreement with the mBJLDA@AE ones.  
The opening in the mBJLDA@PP band gaps is relatively 
larger in the IIB-based compounds (see Fig. \ref{fig:errorsIIA-IIB}), making the ones 
obtained using the cp-PP's very close to the experimental data, while those obtained 
using the csp-PP's are overestimated. 
This trend continues when going to the IIIA-based compounds (see Fig. \ref{fig:errorsIIIB}), 
and in this case the mBJLDA@PP band gaps obtained using csp-PP's and some of the ones 
obtained using the cp-PP's are overestimated.         
 					 
\begin{figure}[H]
	\centering
	\includegraphics[width=0.5\linewidth]{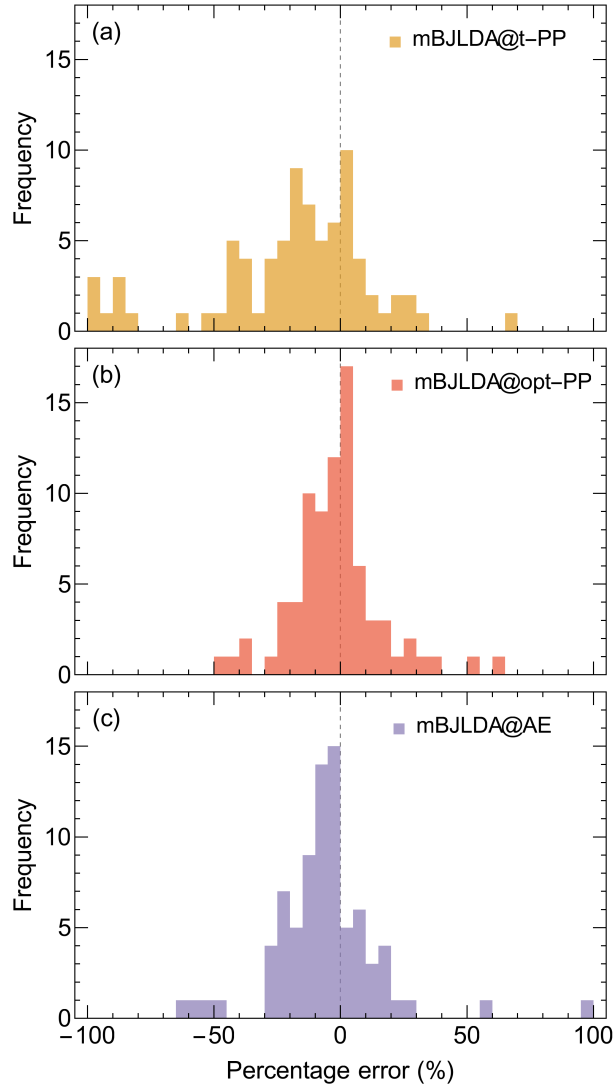}
	\caption{Histograms showing the frequency of the percentage errors in the 
		calculated mBJLDA@PP band gaps using typical PP's (a) and optimal PP's (b). 
		For comparison, in (c) a similar histogram is shown for the percentage 
		errors in reference mBJLDA@AE band gaps \cite{Tran2017,Tran2019,Ye2015}.}
	\label{fig:errors_stats}
\end{figure}

\begin{figure}[H]
	\centering
	\includegraphics[width=1\linewidth]{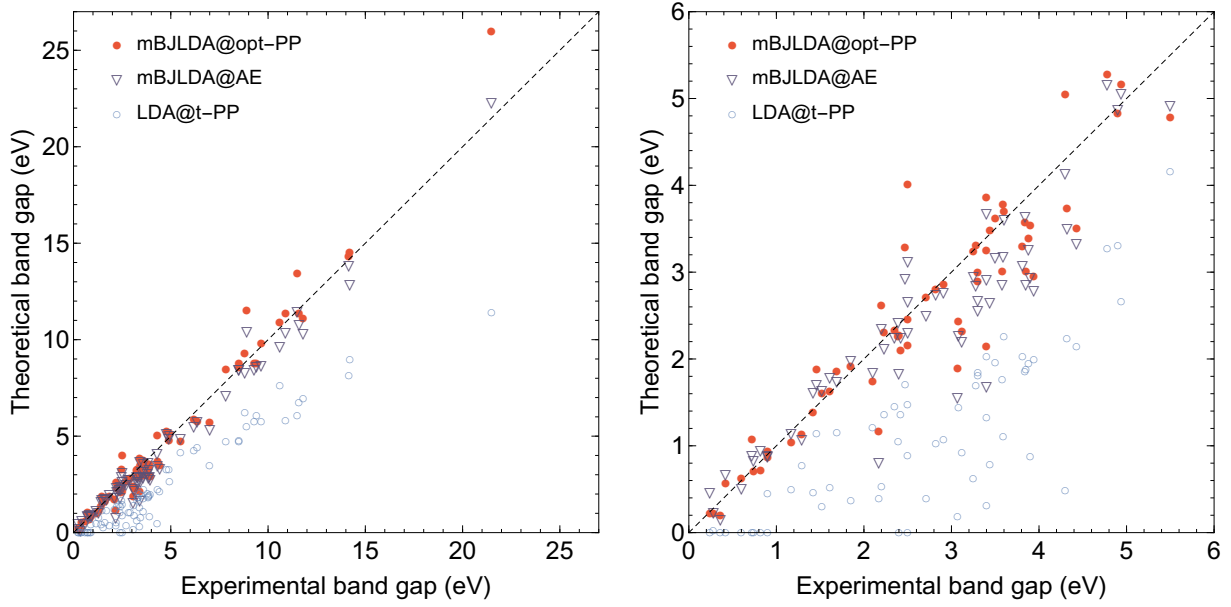}
	\caption{mBJLDA@PP band gaps obtained  using optimal PP's versus experimental 
		band gaps. The right panel is an enlarged view of the left panel focusing on 
		band gaps smaller than 6 eV. For comparison purposes the figure also shows 
		the LDA@t-PP and reference mBJLDA@AE band gaps \cite{Tran2017,Tran2019,Ye2015}.}
	\label{fig:copmpare_expt}
\end{figure}

In order to investigate the overall performance of the mBJLDA@PP approach for band gap 
calculations, we will consider the band gaps obtained by the optimal choices of the 
cations PP's (t-, cd-, cp-, or csp-PP's, that lead to the best overall 
agreement with experiment, as described above), hereafter referred to as opt-PP's. 
In particular,  we considered csp-PP's instead of cp-PP's for Ca, Cu, and Ag because they give
better agreement with experiment (see Sec. \ref{Section:IIA-IIB}). 
In Fig. \ref{fig:errors_stats} we show histograms of the frequency of 
the percentage errors in the calculated mBJLDA@PP band gaps obtained using (a) t-PP's 
and (b) opt-PP's whenever applicable, compared to (c) those of the mBJLDA@AE band gaps. Moreover, in 
Fig. \ref{fig:copmpare_expt} the calculated mBJLDA@opt-PP and mBJLDA@AE band gaps
are compared to experimental data. For more details on the comparison between the mBJLDA@PP 
band gaps and those of mBJLDA@AE and experimental data  we refer the interested readers
to Figs. S1, S2, and S3 in the Supporting Information. 
It can be easily seen from Fig. \ref{fig:errors_stats} that t-PP's tend to lead 
to underestimated band gaps, and this deficiency is highly removed when opt-PP's 
are used. The distribution of percentage errors in mBJLDA@opt-PP band gaps 
become very close to that for the mBJLDA@AE approach. 

One may argue that by treating more core electrons as valence the mBJLDA@PP band gaps 
should approach the corresponding all-electron ones. The better agreement with the mBJLDA@AE 
band gaps achieved by increasing NETV is, generally speaking, in accordance with this argument. 
It is remarkable that an excellent agreement can be achieved by treating only the outermost 
core electrons as valence. The small discrepancies between the mBJLDA@opt-PP and mBJLDA@AE 
band gaps can be attributed to NETV in both approaches and the pseudization of the 
wavefunctions in the former one.     

For a more quantitative analysis we show in Table \ref{table:errors} a summary of 
the statistical analysis of the errors in the calculated band gaps using different 
approaches. Additionally included are the errors in the band gaps obtained with the 
hybrid functional HSE06, taken from Refs. \cite{Tran2017,Borlido2019}.
In addition to the MAE, we also show the mean error, ME, and the mean 
(absolute) percentage error, M(A)PE. The important features to note are the following.
(i) The ME and and MPE in the LDA band gaps, which 
are highly underestimated, are of about $-2$ eV and $-58$\%, 
respectively. 
(ii) These errors are reduced when the mBJLDA@t-PP's
approach is employed. In this case the MPE is about $-19$\%, which reflects the 
tendency of this approach to underestimate the band gaps. The corresponding 
MAE and MAPE values (of 0.85 eV and 26\%) are larger than those obtained 
with mBJLDA@AE calculations (of 0.49 eV and 15\%). 
(iii) When the opt-PP's are used in mBJLDA@PP calculations the errors are substantially 
reduced, with MAE and MAPE of 0.46 eV and bout 12\%, respectively. It can be noted that these 
errors compare very well to those obtained with mBJLDA@AE calculations. 
(iv) The shown ME and MPE values indicate that the mBJLDA@AE approach has more tendency
to underestimate the band gaps than the mBJLDA@opt-PP's method. This is also clear from
Figs. \ref{fig:errors_stats} and \ref{fig:copmpare_expt}. (v) The errors in the 
mBJLDA@opt-PP and mBJLDA@AE approaches compare well with those obtained with the HSE06 method. 

\begin{figure}[H]
	\centering
	\includegraphics[width=0.65\linewidth]{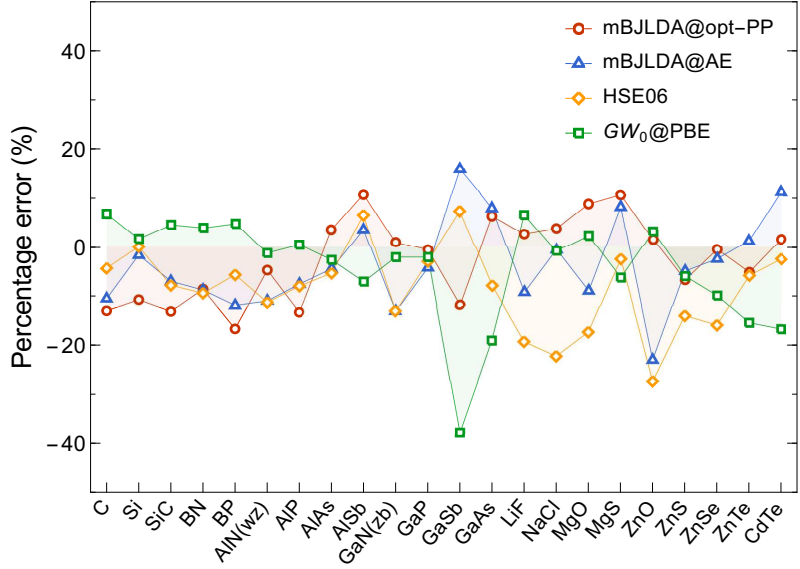}
	\caption{Percentage errors in the $GW_0$@PBE band gaps for selected systems (Ref. \cite{Jiang2016}) 
	together with the corresponding errors in the mBJLDA@opt-PP, mBJLDA@AE (Ref. \cite{Tran2017}) and 
	HSE06 band gaps (Ref. \cite{Tran2017}). The points are connected as a guide to the eye.}
    \label{fig:errors_GW}
\end{figure}

Finally, we compare the mBJLDA band gaps with those of the $GW$ approximation, which is 
the most formal and rigorous approach for band structure calculations. Before doing that, 
it worth noting that there are 
several technical details that have significant effects on the single-shot $GW$ 
(or $G_0W_0$) band gaps (see for example Ref. \cite{Golze2019}). Moreover, in 
addition to $G_0W_0$, partially \cite{Jiang2016,Shishkin2007} and fully self-consistent 
\cite{Grumet2018,Kutepov2017} $GW$ approaches have been developed, and they provide 
different levels of agreement with experiment \cite{Golze2019}. Therefore, a comprehensive 
comparison between mBJLDA and $GW$ band gaps is not a simple task. As a representative, we 
will consider the $GW_0$@PBE band gaps reported by Jiang and Blaha \cite{Jiang2016}, for 
a much shorter list of compounds. 
The percentage errors in these $GW_0$ band gaps are shown in Fig. \ref{fig:errors_GW}, 
together with the corresponding ones in the mBJLDA@opt-PP, mBJLDA@AE and HSE06 band gaps. 
It can be noted that all these methods generally provide very 
good agreement with experiment. The $GW_0$ band gaps have the 
lowest ME (MAE) of $-$0.02 (0.21) eV, compared to $-$0.04 (0.25) eV using mBJLDA@opt-PP, 
$-$0.24 (0.32) eV using mBJLDA@AE, and $-$0.49	(0.51) eV using HSE06. On the other hand, 
the calculated MAPE values in the $GW_0$, mBJLDA@opt-PP, mBJLDA@AE and HSE06 band gaps 
are respectively of 7.3\%, 7.1\%, 8.1\% and 9.9\%. 

\begin{table}[H]
	\centering
	\begin{tabular}{lSSSSS} 
		\toprule
		& \multicolumn{1}{c}{LDA} & \multicolumn{3}{c}{mBJLDA} & \multicolumn{1}{c}{HSE06}  \\
		\cmidrule{3-5}
		& & \multicolumn{1}{c}{@t-PP} & \multicolumn{1}{c}{@opt-PP} & \multicolumn{1}{c}{@AE} & \\ 
		\midrule
		ME (eV)  & -2.15  & -0.41 &  0.01 & -0.33 & -0.69 \\
		MAE (eV) &  2.15  &  0.85 &  0.46 &  0.49 &  0.82 \\
		MPE (\%) & -57.63 &-18.82 & -1.63 & -6.01 & -7.57 \\
		MAPE (\%) & 57.63 & 26.31 & 12.41 & 15.20 & 17.15 \\
		\bottomrule
	\end{tabular}
	\caption{Mean (absolute) error, M(A)E, and mean (absolute) percentage error, M(A)PE, in 
		the band gaps using different approaches, with respect to experiment. The mBJLDA@AE 
			and HSE06 band gaps are taken from Refs. \cite{Tran2017,Tran2019,Ye2015} and Refs. 
			\cite{Tran2017,Borlido2019}, respectively. For more details see Table S1 in the 
			Supporting Information.}
	\label{table:errors}
\end{table}

\section{Conclusions}

In this work we performed mBJLDA calculations employing a norm-conserving 
pseudopotential plane-waves approach (mBJLDA@PP) for a test set of 83 solids, 
representing a wide range of semiconductors and insulators. We investigated 
the effect of the number of electrons treated as valence in the PP's of the cations. 
The obtained band gaps are discussed in comparison with those obtained using 
all-electron mBJLDA (mBJLDA@AE), $GW_0$, HSE06, and experimental 
data. The main conclusions based on the results obtained for the above considered 
systems can be summarized as follows.

\begin{enumerate}
	\item For the group IA-based compounds, the use of typical pseudopotentials 
	(t-PP's), where valence and semicore states are pseudized, leads to band gaps 
	that are in very good agreement with the mBJLDA@AE ones and experiment.
	
	\item For the group IIA-based compounds, the best overall agreement between the 
	mBJLDA@PP and mBJLDA@AE band gaps is obtained by using t-PP's. The effects of including 
	uppermost core states as valence are quite small. However, including the uppermost $s$ 
	and $p$ (in Ca PP) and $d$ (in the Sr and Ba PP's) core electrons as valence leads to 
	mBJLDA@PP band gaps in better agreement with experiment. These PP's are referred to 
	as csp-PP's (cd-PP's).
	
	\item For groups IB-, IIB-, and IIIA-based compounds the best overall 
	agreement with the mBJLDA@AE band gaps is obtained when t-PP's (for 
	light elements, namely Be, Mg, B and Al) and cp-PP's (for the heavier elements) 
	are employed. In the case of Cu- and Ag-based compounds, a better agreement 
	with experiment is achieved when csp-PP's are employed.
	
	\item For transition-metal based compounds the best overall agreement 
	with the mBJLDA@AE and experimental band gaps is obtained by using cp-PP's.	
	
	\item For group IVA elemental solids and compounds, rare gas solids, and Sb$_2$Te$_3$ 
	the best overall agreement with the mBJLDA@AE and experimental band gaps 
	are obtained by using t-PP's.
	
	\item The relative opening of the mBJLDA@PP band gaps by including the uppermost core 
	states as valence increases by going from IB- to IIB- to IIIA-based compounds.
	
	\item The overall accuracy of the mBJLDA@PP approach when the above optimal PP's are employed 
	is very close to that of $GW_0$ and slightly better than that of the mBJLDA@AE and HSE06 methods.
	
\end{enumerate}	  

\section*{Acknowledgments} 
The authors would like to thank Mohammed Abu-Jafar for fruitful discussion. The first author would like to
thank Yarmouk University for providing the computational resources, and Prince Sultan University for their support. 

\bibliographystyle{elsarticle-num}

\end{document}